\documentclass[twocolumn]{aastex6}

\def \Msol{{\rm M}_{\odot}}
\def \Lsol{{\rm L}_{\odot}}

\def \logm{\log(M/\Msol)}
\def \lya{Ly$\alpha$}

\def \h2{{\rm H_{2}}}
\def \oabund{12+\log({\rm O/H})}

\def \siii{Si{\scriptsize ~II}}

\def \siiv{Si{\scriptsize ~IV}}
\def \cii{[C{\scriptsize ~II}]}

\def \oii{[O{\scriptsize ~II}]}
\def \oiii{[O{\scriptsize ~III}]}

\def \nv{N{\tiny\,V}}

\def \LIR{L_{{\rm FIR}}}
\def \LUV{L_{{\rm 1600}}}

\def \LCII{L_{{\rm CII}}}
\def \IRXB{IRX$-\beta$}
\def \dn4000{D_{{\rm n}}(4000) }

\defcitealias{CAPAK15}{C15}

\begin{document}


\title{Dust Properties of \cii$-$detected $z\sim5.5$ galaxies: new HST/WFC3 near-IR Observations}


\author{
Ivana Bari\v si\' c$^{1,2}$,
Andreas L. Faisst$^{3}$,
Peter L. Capak$^{3}$,
Riccardo Pavesi$^{4}$,
Dominik A. Riechers$^{4}$,
Nick Z. Scoville$^{5}$,
Kevin C. Cooke$^{6}$,
Jeyhan S. Kartaltepe$^{6}$,
Caitlin M. Casey$^{7}$,
Vernesa Smol\v ci\' c$^{1}$
}





\affil{$^1$Department of Physics, Faculty of Science, University of Zagreb,  Bijeni\v{c}ka cesta 32, 10000  Zagreb, Croatia}
\affil{$^2$Max-Planck Institut f\"ur Astronomie, K\"onigstuhl 17, D-69117, Heidelberg, Germany}
\affil{$^3$Infrared Processing and Analysis Center, California Institute of Technology, Pasadena, CA 91125, USA}
\affil{$^4$Department of Astronomy, Cornell University, Space Sciences Building, Ithaca, NY 14853, USA}
\affil{$^5$Cahill Center for Astronomy and Astrophysics, California Institute of Technology, Pasadena, CA 91125, USA}
\affil{$^6$School of Physics and Astronomy, Rochester Institute of Technology, 84 Lomb Memorial Drive, Rochester, NY 14623, USA}
\affil{$^7$Department of Astronomy, The University of Texas at Austin, 2515 Speedway Blvd Stop C1400, Austin, TX 78712, USA}


\email{barisic@mpia.de}

\begin{abstract}
	We examine the rest-frame ultra-violet (UV) properties of 10 \cii$\lambda158\,{\rm \mu m}$$-$detected galaxies at $z\sim5.5$ in COSMOS using new HST/WFC3 near-infrared imaging. Together with pre-existing $158\,{\rm \mu m}-$continuum and \cii~line measurements by ALMA, we study their dust attenuation properties on the \IRXB~diagram, which connects the total dust emission ($\propto$ ${\rm IRX}=\log(\LIR/\LUV)$) to the line-of-sight dust column ($\propto\beta$).
	We find systematically bluer UV continuum spectral slopes ($\beta$) compared to previous low-resolution ground-based measurements, which relieves some of the tension between models of dust attenuation and observations at high redshifts. While most of the galaxies are consistent with local starburst or Small Magellanic cloud like dust properties, we find galaxies with low IRX values and a large range in $\beta$ that cannot be explained by models of a uniform dust distribution well mixed with stars.
	A stacking analysis of Keck/DEIMOS optical spectra indicates that these galaxies are metal-poor with young stellar populations which could significantly alter their spatial dust distribution.

\end{abstract}




\section{Introduction}

	Galaxies at $z>4$ likely have different inter-stellar medium (ISM) properties than galaxies in the local Universe \citep[e.g.,][]{CARILLI13}.
	For example, studies of emission line ratios of high redshift ($z > 4$) galaxies by Spitzer suggest that early galaxies grow at a fast pace, with high specific star-formation rates (SFR) and stellar mass assembly time scales of only a couple of $100\,{\rm Myrs}$ \citep[][]{DEBARROS14,FAISST16a,JIANG16}. A substantially higher gas fraction as well as changes in the ionization parameter of stellar radiation indicated by the \oiii/\oii~line ratio are also observed, indicative of a different H$_2$ region configuration \citep[][]{STEIDEL14,GENZEL15,SILVERMAN15,SCOVILLE16,FAISST16c,MASTERS16}.
	
	So far, direct studies of the ISM at $z>4$ have been difficult due to the lack of sensitivity in the far-infrared (FIR) and sub-millimeter (sub-mm) wavelengths where the key ISM diagnostics are present.  
	With the \textit{Atacama Large Millimeter/Sub-millimeter Array} (ALMA) it is now possible to study the ISM of galaxies at very high redshifts via the measurement of the \cii~$158\,{\rm \mu m}$ emission and the FIR continuum. Some of the first studies of $z>4$ galaxies with ALMA find significantly lower dust continuum emission than those found in local galaxies \citep[][]{RIECHERS14,CAPAK15,MAIOLINO15,SCHAERER15,WATSON15,BOUWENS16,CARILLI16,DUNLOP16a,KNUDSEN16}.
	
	Studying a sample of $10$ Lyman Break Galaxies (LBGs) at $5 < z < 6$, \citet{CAPAK15} (hereafter \citetalias{CAPAK15}) find the ratio of infrared to ultraviolet light ($\LIR/\LUV~=~{\rm IRX}$) at a given UV continuum spectral slope ($\beta$) in the so-called \IRXB~relation to be significantly lower than what was expected from dust attenuation models \citep[e.g.,][]{CHARLOTFALL00} or even in local low-metallicity galaxies like the Small Magellanic Cloud \citep[SMC,][]{PETTINI98}. Furthermore, there are galaxies with exceptionally red $\beta$ slopes at low IRX values, which are difficult to physically explain \citep[][]{CHARLOTFALL00}. These findings are curious and, if confirmed, would imply a substantial evolution of ISM properties in the first $1\,{\rm billion}$ years of the Universe.
	These measurements leave significant uncertainty in what is physically occuring.
	First, the FIR luminosities are derived from a single ALMA continuum point at rest-frame $158\,{\rm \mu m}$, which means significant assumptions on the shape of the FIR spectral energy distribution (SED) must be made.
	Second, the $\beta$ measurements are derived from low signal-to-noise (S/N) ground-based data, which results in significant systematic uncertainty \citep[e.g.,][]{DUNLOP12,ROGERS13}.
	
	In this letter, we present deep \textit{Hubble Space Telescope} (HST) near-IR imaging to provide improved measurements of $\beta$ and the UV luminosity ($\LUV$) for the 10 originally targeted, two serendipitous sources, and two sub-components of one of the galaxies that are resolved by HST at $z\sim5.5$ presented in \citetalias{CAPAK15} (Section~\ref{sec:data}). We carefully investigate potential systematics in the derivation of these parameters by performing extensive simulations to eventually pinpoint the location of these galaxies on the \IRXB~plane (Sections~\ref{sec:measurement}~and~\ref{sec:results}).
	A companion paper (Faisst et al., in prep) presents an additional analysis of the dust temperatures and far-infrared properties of these galaxies given a sample of local galaxies.
	
	We assume a flat cosmology with $\Omega_{\Lambda,0} = 0.7$, $\Omega_{m,0} = 0.3$, and $h = 0.7$. Magnitudes are quoted in AB \citep{OKE74}.

\section{Data}\label{sec:data}

\subsection{Sample selection and auxiliary data}

\begin{figure*}
\centering
\includegraphics[width=\linewidth]{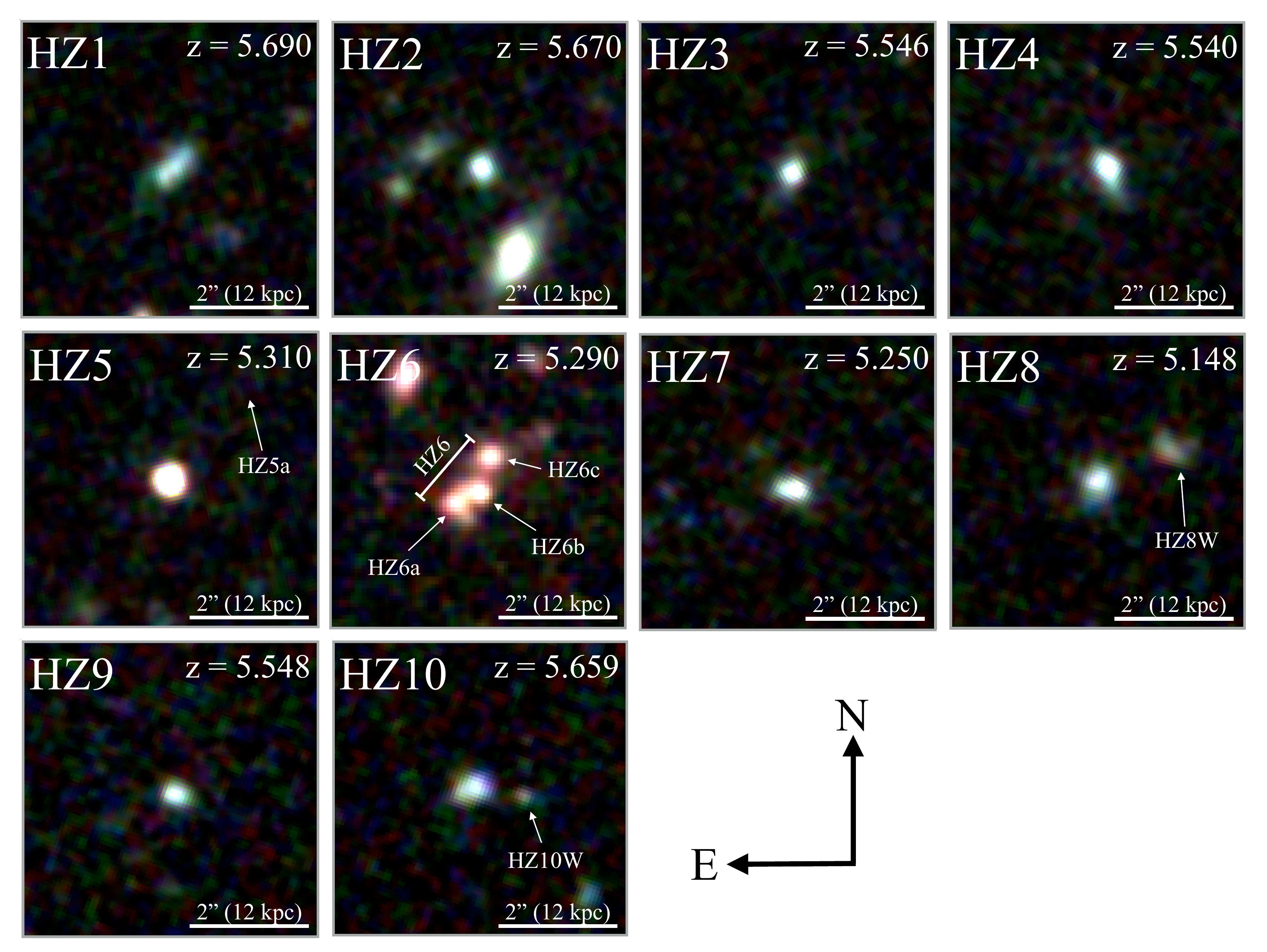} 
\caption{HST/WFC3 color (F105W (blue), F125W (green), F160W (red)) cut-outs from the 10 pointings. The two serendipitous \cii~detections HZ5a and HZ8W and other sub-components are indicated by white arrows.
\label{ds9}}
\end{figure*}

	The sample that is used in this work is based on \citetalias{CAPAK15} and we describe its properties in brief in the following.
	The sample consists of 9 star-forming LBGs at $1-4$ times the characteristic luminosity ($L^{*}$) and one low luminosity quasar {\textit{HZ5}, see \citet{CAPAK11}} observed at $5 < z < 6$ in the 2 square-degree field of the \textit{Cosmic Evolution Survey} \citep[COSMOS,][]{SCOVILLE07}\footnote{\url{http://cosmos.astro.caltech.edu}}, which provides a multitude of photometric data \citep[][]{LAIGLE16}.
	Note that we will not use HZ5 in the following analysis of this work. The sample is initially selected by a range of techniques and designed to be representative of z = 5 - 6 galaxies in many quantities including SFR, stellar mass, and metallicity. The ten objects are followed up with the \textit{Deep Extragalactic Imaging Multi-Object Spectrograph} \citep[DEIMOS,][]{FABER03} to spectroscopically confirm their redshifts via the rest-frame UV absorption features. Out of ten galaxies, six show a strong Ly$\alpha$ emission while three show only weak emission. No Ly$\alpha$ was detected in \textit{HZ3}. The galaxies cover a large range in SFRs and stellar masses (see \citetalias{CAPAK15}) and the UV absorption features suggest gas-phase metallicities of $\oabund = 8.0-8.5$, representative of the average metallicity of $\logm\sim10$ galaxies at $z\sim5$ \citep[][]{FAISST16b}.
	The galaxies were observed with ALMA\footnote{ALMA Cycle 1, ADS/JAO.ALMA\#2012.1.00523.S, PI: P. L. Capak. See \citetalias{CAPAK15} for a detailed description of this data.} targeting the wavelength around rest-frame $158\,{\rm \mu m}$ including the FIR continuum and the \cii~emission line \citep{CAPAK15}. All galaxies were detected in \cii~and four of them have detected FIR continuum. In addition, two serendipitously detected \cii~emitters (\textit{HZ5a} and \textit{HZ8W}) were found at the same \cii~redshift of \textit{HZ5} and \textit{HZ8} at a projected distance of $10-15\,{\rm kpc}$ and \textit{HZ6} breaks into three sub-components for a total of $14$ objects (Figure~\ref{ds9}).

	\subsection{New HST/WFC3 observations}
	
	To study the rest-frame UV properties of these galaxies, we have obtained HST Wide Field Camera 3 (WFC3) imaging between March and July 2015  (HST Cycle 22, ID 13641, PI: P. L. Capak) in F105W, F125W, and F160W corresponding to $1.05\,{\rm \mu m}$, $1.25\,{\rm \mu m}$, and $1.54\,{\rm \mu m}$, respectively. The ten galaxies were observed in ten separate pointings for one orbit per filter reaching $26.2$, $26.4$, and $25.9\,{\rm AB}$ ($5\sigma$ in $1\arcsec$ aperture). Note that \textit{HZ6} was observed in another HST program (HST Cycle 21, ID 13384, PI: D. Riechers) to which we added F105W.
	The standard HST data reduction procedure as explained in the HST Data Handbook\footnote{\url{http://www.stsci.edu/hst/wfc3/analysis}} was used to convert the raw data frames into the final flat-fielded and flux-calibrated science products. Thereby we used the Pyraf/STSDAS \texttt{calfw3} task to create a bad pixel array and to perform the bias and dark current subtraction of the read-outs.  
	Geometric distortions have been corrected for using the latest distortion solutions (March 22 2012) provided by the STScI/WFC3 web-page and the exposures are joined using the Pyraf/STSDAS \texttt{AstroDrizzle} task.
	We use the HST/ACS F814W images available on the COSMOS field for the astrometric alignment of the frames to reach sub-arcsecond precision (Cooke et al. in prep).
	Figure~\ref{ds9} shows HST/WFC3 3$-$color cut-outs of each of the 10 pointings.

\section{Measurements \& Simulations}\label{sec:measurement}

	The UV continuum slope $\beta$ for star forming galaxies at high-z is dominated by the line-of-sight dust column density, which attenuates the UV light of the star forming regions\footnote{Note that $\beta$ also depends on internal properties of a galaxy such as the age and metallicity of a stellar population as well as its star-formation history. However, these affect less the overall $\beta$ at high redshifts as compared to dust \citep[see figure~13 in][]{BOUWENS12}}. 
	It is therefore a valid tool to study the ISM column density of galaxies \citep[][]{MEURER99,KONG04,CASEY14}. Furthermore, $\beta$ is the only estimate of the dust content available for the vast majority of galaxies at $z>3$ due to the relative insensitivity of contemporary optical and IR surveys. The $\beta$ can be combined with IRX, which is sensitive to the total amount of dust, to form the \IRXB~relation which can be used to infer the evolution of ISM properties.
	
	The measurement of $\beta$ at high redshifts is not trivial due to the scarcity of photometric bands covering the rest-frame UV part of the galaxies' SEDs and low sensitivity from the ground. Furthermore, the measurement of $\beta$ is particularly sensitive to the signal-to-noise of photometry and systematics in flux measurements, such as confusion from neighboring sources, background subtraction, and surface brightness effects \citep[e.g.,][]{FINKELSTEIN12,DUNLOP12,ROGERS13,BOUWENS14}.
	High S/N data accompanied by simulations are therefore essential to correct for these effects. In the following sections we describe the measurement of the $\beta$ and the methods used to access and correct systematic uncertainties.

\subsection{Flux extraction and measurement of $\beta$}\label{sec:betameasurement}

\begin{figure*}
\centering
\includegraphics[width=0.8\linewidth, angle=0]{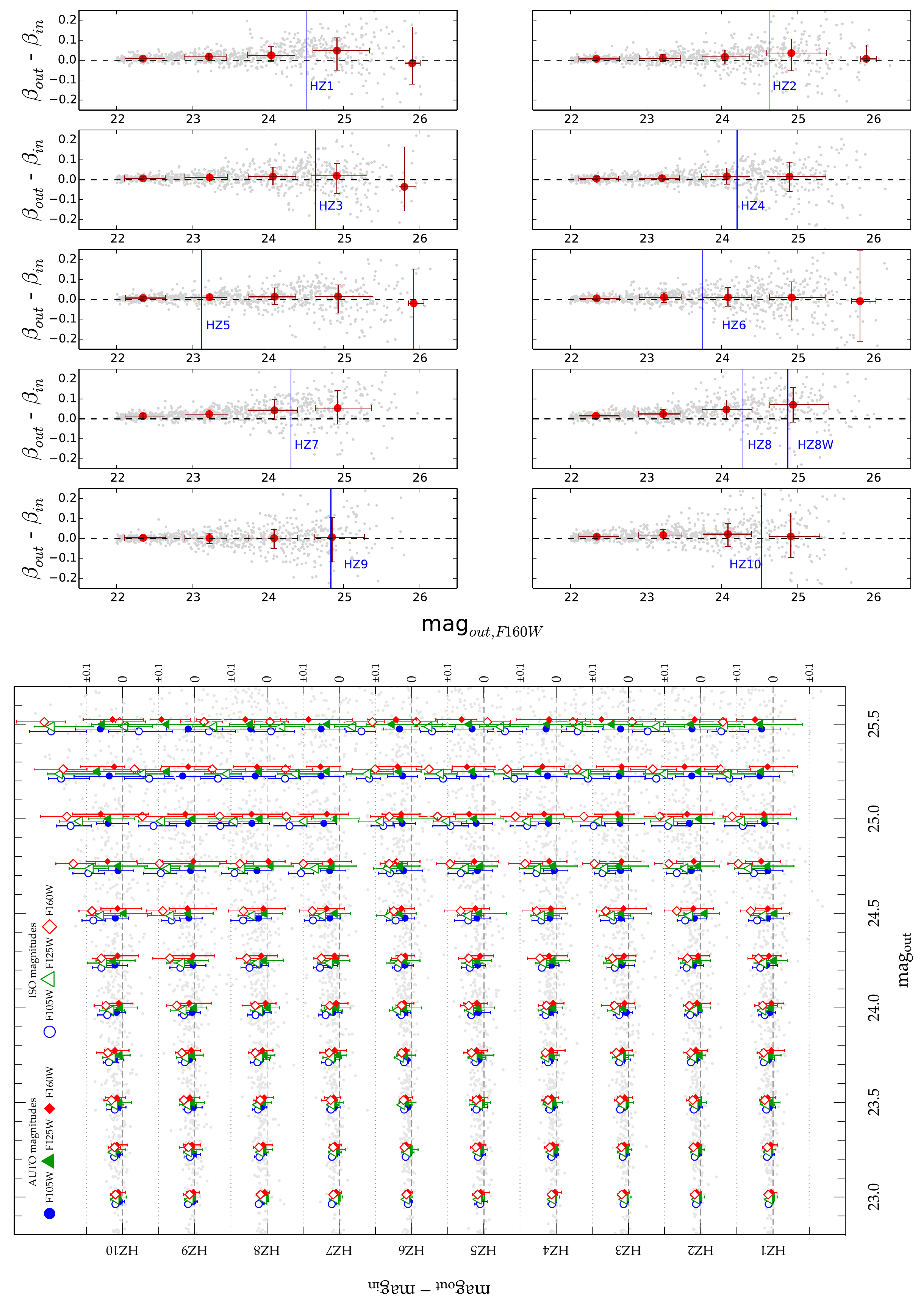}
\caption{Results from our simulations.
\textit{Top:} Recovering error of $\beta$ as a function of output F160W magnitude for each HST pointing (gray for single galaxies, red in bins of magnitude). The simulation suggests no significant biases.
\textit{Bottom:} Recovering error in F105W (blue dots), F125W (green triangles), and F160W (red diamonds) magnitudes for the ten HST pointings (filled: \texttt{AUTO} magnitude; open: \texttt{ISO} magnitudes). \texttt{ISO} magnitudes deviate systematically from the true by more than $0.1\,{\rm mag}$ at $>24.5\,{\rm AB}$ in comparison to \texttt{AUTO} magnitudes ($<0.1\,{\rm mag}$ up to $25.5\,{\rm AB}$).
\label{mag}}
\end{figure*}

	We use \textit{Source Extractor} \citep[version 2.5.0,][]{BERTIN96} to measure the fluxes of the galaxies in F105W, F125W, and F160W. For each of the galaxies, we manually tune the \textit{Source Extractor}'s input parameters  in order to optimize the de-blending, object detection, and background subtraction.
	The fluxes are measured in different aperture types including \texttt{ISO}, \texttt{AUTO}, as well as three manual apertures with radii $r=0.8\arcsec$, $0.9\arcsec$, and $1.0\arcsec$. The isophotal flux threshold was set to $2\sigma$ relative to the weight map and the \texttt{AUTO} apertures use a minimum radius of $3.5$ pixels ($0.45\arcsec$) for low S/N sources.
	We choose the Kron-factor (2.5) and the minimum radius (3.5px) to be the same for all pointings and wavelengths. From our realistic and extensive simulations (Figure~\ref{mag} , described in Section~\ref{sec:simulations} ), we find that we can recover the \texttt{AUTO} magnitudes better than 0.05 mag down to 25AB magnitude. The simulations and visual inspection of the output photometry show that the \texttt{AUTO} magnitudes have the least systematic biases compared to other methods of measuring the photometry of the galaxies.	
	
	In addition to the these main measurements, the near-IR HST photometry of the sub-components of \textit{HZ6} are extracted using \texttt{GALFIT} \citep[version 3.0.5,][]{CHIEN10} and additionally verified by manual apertures (agreement to better than $0.1\,{\rm mag}$). The FIR \cii~line emission and $158\,{\rm \mu m}$ continuum fluxes at the positions of the three sub-components are fitted by Gaussians with the size of the ALMA beam ($\sim 0.5\arcsec$). {\textit{HZ6a} and \textit{HZ6b}} are significantly blended with ALMA and therefore the relative flux extraction is uncertain by up to a factor of three depending on the weighting.
	 	
	The UV spectral slopes of the galaxies are measured directly from the F105W, F125W, and F160W photometric bands, which correspond to rest-frame $1600\,{\rm \AA}$ to $2400\,{\rm \AA}$ at $z\sim5.5$. We note that the effective center wavelength of the filters changes with the power-law function $f_{\lambda} \propto \lambda^\beta$ and must be accounted for in the measurement of $\beta$, since the filter transmission curves have a finite width and a calibration on the AB system assumes $\beta=-2$.
	We therefore use a forward-modeling approach to measure $\beta$ for our galaxies. From a tightly spaced grid in intrinsic spectral slopes ($\beta_{in}$) and intercepts ($C$), we create power-law SEDs according to $\log(f_{\lambda}) = \beta_{in}\,\times\,\log(\lambda)+C$, which we convolve with the three filter transmission curves for F105W, F125W, and F160W, respectively. The best $\beta_{in}$ is then found via $\chi^2~-$~minimization from the comparison of the previously obtained model fluxes and the measured fluxes of the galaxies.  We obtain estimates of the measurement uncertainties on $\beta$ by repeating the above steps and thereby varying the observed fluxes within their errors (assuming a Gaussian error distribution).
	In addition we add a 0.007 magnitude error coming from the uncertainties in the WFC3-IR detector chain calibration (S. Deustua, WFC3 team, private comm).
	The extracted photometry of the galaxies as well as the final $\beta$ and $\LUV$ with $1\sigma$ uncertainties are listed in Table~\ref{tab:photbeta}.

\begin{deluxetable*}{lcccccccc}
\tabletypesize{\scriptsize}
\tablecaption{Summary of HST ultra-violet and ALMA FIR measurements for the \citetalias{CAPAK15} sample\label{tab:photbeta}}
\tablewidth{0pt}
\tablehead{
\colhead{} & \colhead{} & \multicolumn{5}{c}{HST based measurements (this work)} & \multicolumn{2}{c}{ALMA FIR measurements$^{a}$}\\[-0.2cm]
\colhead{} & \colhead{} & \multicolumn{5}{c}{------------------------------------------------------------------------------------------------} & \multicolumn{2}{c}{-----------------------------------------}\\[-0.1cm]
\colhead{ID} & \colhead{redshift} & \colhead{F105W$^{h}$} & \colhead{F125W$^{h}$} & \colhead{F160W$^{h}$} & \colhead{$\beta$} & \colhead{$\log(\LUV/\Lsol)$} & \colhead{$\log(\LIR/\Lsol)$} & \colhead{$\log(\LCII/\Lsol)$}
}
\startdata
HZ1 & 5.690 & 24.53 $\pm$ 0.04 & 24.31 $\pm$ 0.03 & 24.52 $\pm$ 0.05 & -1.92$_{-0.11}^{+0.14}$ & 11.21 $\pm$ 0.01 & $\textless$ 10.32 & 8.40 $\pm$ 0.32 \\
HZ2 & 5.670 & 24.65 $\pm$ 0.03 & 24.50 $\pm$ 0.03 & 24.63 $\pm$ 0.05 & -1.82$_{-0.10}^{+0.10}$ & 11.15 $\pm$ 0.01 & $\textless$ 10.30 & 8.56 $\pm$ 0.41  \\
HZ3 & 5.546 & 24.74 $\pm$ 0.04 & 24.57 $\pm$ 0.03 & 24.63 $\pm$ 0.04 & -1.72$_{-0.15}^{+0.12}$ & 11.08 $\pm$ 0.01 & $\textless$ 10.53 & 8.67 $\pm$ 0.28 \\
HZ4 & 5.540 & 24.17 $\pm$ 0.02 & 24.21 $\pm$ 0.02 & 24.20 $\pm$ 0.03 & -2.06$_{-0.15}^{+0.13}$ & 11.28 $\pm$ 0.01 & 11.13 $\pm$ 0.54 & 8.98 $\pm$ 0.22 \\
HZ5 & 5.310 & 23.48 $\pm$ 0.01 & 23.53 $\pm$ 0.01 & 23.12 $\pm$ 0.01 & -1.01$_{-0.12}^{+0.06}$ & 11.45 $\pm$ 0.01 & $\textless$ 10.30 & $\textless$ 7.20\\
HZ5a & 5.310 & $>26.24$$^b$ & $>26.35$$^b$ & $>25.89$$^b$ & \nodata  & $<10.37$$^b$ &  $\textless$ 10.30 & 8.15 $\pm$ 0.27\\
%
HZ6 (total)$^c$ & 5.290 & 23.62 $\pm$ 0.03 & 23.45 $\pm$ 0.04 & 23.29 $\pm$ 0.03 & -1.14$_{-0.14}^{+0.12}$ & 11.47 $\pm$ 0.10 & $11.13\pm0.23$$^c$ & $9.23\pm0.04$$^{c}$\\
$\cdot\;\;$HZ6a$^d$ &  & 25.28 $\pm$ 0.36 & 24.90 $\pm$ 0.30 & 24.67 $\pm$ 0.25 & -0.59$_{-1.12}^{+1.05}$ & 11.11 $\pm$ 0.07 & $10.26\pm0.23$$^d$ & $8.32\pm0.06$$^{d,e}$\\
$\cdot\;\;$HZ6b$^d$ &  & 24.77 $\pm$ 0.28 & 24.65 $\pm$ 0.33 & 24.58 $\pm$ 0.37 & -1.50$_{-1.22}^{+1.05}$ & 11.00 $\pm$ 0.07 & $10.87\pm0.23$$^d$ & $8.81\pm0.02$$^{d,e}$\\
$\cdot\;\;$HZ6c$^d$ &  & 24.53 $\pm$ 0.16 & 24.43 $\pm$ 0.13 & 24.25 $\pm$ 0.10 & -1.30$_{-0.37}^{+0.51}$ & 10.81 $\pm$ 0.07 & $10.79\pm0.23$$^d$ & $8.65\pm0.03$$^{d}$\\
HZ7 & 5.250 & 24.56 $\pm$ 0.04 & 24.42 $\pm$ 0.03 & 24.30 $\pm$ 0.03 & -1.39$_{-0.17}^{+0.15}$ & 11.05 $\pm$ 0.02 & $\textless$ 10.35 & 8.74 $\pm$ 0.24 \\
HZ8 & 5.148 & 24.51 $\pm$ 0.04 & 24.37 $\pm$ 0.04 & 24.28 $\pm$ 0.04 & -1.42$_{-0.18}^{+0.19}$ & 11.04 $\pm$ 0.02 & $\textless$ 10.26 & 8.41 $\pm$ 0.18 \\
HZ8W & 5.148 & 25.68 $\pm$ 0.11 & 25.14 $\pm$ 0.07 & 24.88 $\pm$ 0.08 & -0.10$_{-0.29}^{+0.29}$ & 10.57 $\pm$ 0.04 & $\textless$ 10.26 & 8.31 $\pm$ 0.23  \\
HZ9 & 5.548 & 24.99 $\pm$ 0.06 & 24.91 $\pm$ 0.05 & 24.83 $\pm$ 0.07 & -1.59$_{-0.23}^{+0.22}$ & 10.95 $\pm$ 0.02 & 11.54 $\pm$ 0.19  & 9.21 $\pm$ 0.09  \\
HZ10 & 5.659 & 24.56 $\pm$ 0.05 & 24.48 $\pm$ 0.05 & 24.52 $\pm$ 0.06 & -1.92$_{-0.17}^{+0.24}$ & 11.14 $\pm$ 0.02 & 11.94 $\pm$ 0.08  & (9.60 $\pm$ 0.13)$^f$ \\
  $\cdot\;\;$HZ10W$^g$ &  & 26.83 $\pm$ 0.16 & 26.86 $\pm$ 0.13 & 26.63 $\pm$ 0.15 & -1.47$_{-0.44}^{+0.77}$ & 10.23 $\pm$ 0.05 & (11.64 $\pm$ 0.08)$^g$ & (8.93 $\pm$ 0.13)$^g$ \\
\enddata
\tablenotetext{a}{Measurements by \citetalias{CAPAK15} unless noted differently. The FIR luminosity is integrated between $3\mu m$ and $1100\mu m$.}
\tablenotetext{b}{$5\,\sigma$ limits in $1\arcsec$ aperture.}
\tablenotetext{c}{Integrated photometry of all the sub-components. We use the ALMA data:  ADS/JAO.ALMA\# 2011.0.00064.S  \citep{RIECHERS14, PAVESI16}.}
\tablenotetext{d}{The photometry for the sub-components of HZ6 is measured by \texttt{GALFIT}. The ALMA properties are re-extracted using the F160W data as spatial prior.}
\tablenotetext{e}{The quoted uncertainties are formal fitting uncertainties. The actual uncertainties for \textit{HZ6a} and \textit{HZ6b} are larger (up to $0.2\,{\rm dec}$) due to severe blending.}
\tablenotetext{f}{Updated value from \citet{PAVESI16}.}
\tablenotetext{g}{Photometry measured in a $0.5\arcsec$ aperture. We assume half the FIR and \cii~luminosity of \textit{HZ10}.}
\tablenotetext{h}{{The quoted errors are measurement noise. For computing $\beta$ we have added 0.007 mag of systematic uncertainty due to the WFC3-IR detector chain calibration (S. Deustua, WFC3 team, private comm).  The systematic noise due to the photometry method is included in the simulations.  The conversion of WFC3-IR fluxes to the absolute frame is uncertain at the 2.1$\%$ level, but is dominated by the uncertainty of the standard star system which does not affect the measured colors. }}

\end{deluxetable*}

\subsection{Simulations}\label{sec:simulations}

	We now assess possible systematic uncertainties of our $\beta$ measurements arising from biases in the extraction of the photometry. For this, we create model galaxies, which are placed onto the original HST pointings and re-extracted with the same parameters used for the real galaxies.
	Following our approach in Section~\ref{sec:betameasurement}, we create $1,000$ random model SEDs with $-3 < \beta < 2$ and $22 < {\rm mag_{F160W}} < 26$ from which we obtain F105W, F125W, and F160W total fluxes by convolving with the respective filter transmission curves. The actual model galaxies are parametrized by a 2-dimensional Gaussian light distribution scaled to their total fluxes in each filter and with half-light sizes equal to the median measured for the real galaxies in the respective filters ($\sim0.17\arcsec$).
	The model galaxies are then placed randomly in empty spots on each of the $10$ HST calibrated pointings. For this we stack the segmentation images in all three filters as returned by \textit{Source Extractor} and require a box of $20\times20$ pixels of clean sky background around the coordinate at which the model galaxies is inserted onto the frame.
	The model galaxies are re-extracted and analyzed in the same way as the real galaxies in order to obtain their output fluxes and spectral slope ($\beta_{out}$).
	
	Figure~\ref{mag} shows the results of our simulations. Each of the $10$ panels in the top figure correspond to one HST pointing and shows the difference $\beta_{out}-\beta_{in}$ as a function of output F160W magnitude. The observed magnitude of the $10$ galaxies in each frame including the two serendipitous \cii~detections~is indicated with blue lines. The increase in scatter for fainter magnitudes (i.e., lower S/N) is clearly visible. For brighter magnitudes we find a fast convergence to $\beta_{out}-\beta_{in}=0$ as expected. In general, we find systematic offsets in our $\beta$ measurements of less than $\Delta\beta=0.05$ at the observed magnitudes of our galaxies, which is several factors less than the actual measurement uncertainties. We therefore conclude that our $\beta$ measurements are robust and not significantly biased. We also note that the scatter from the true $\beta$ seen in the simulation is consistent with the $1\sigma$ uncertainty we derive for the $\beta$ measurements (see Table~\ref{tab:photbeta}), which suggests that the dominant factor of uncertainty is photometric and a power-law approximation for the galaxies' SEDs is reasonable.
	We do not find significant biases ($<0.05\,{\rm mag}$ up to $25.5\,{\rm AB}$) in recovering the \texttt{AUTO} magnitudes (lower panel in Figure~\ref{mag}). Finally, we note that our simulations show that the photometry computed in \texttt{ISO} and manual (1") apertures are systematically biased at $>24.5\,{\rm AB}$ by $0.1\,{\rm mag}$ and $0.15\,{\rm mag}$, respectively, due to surface brightness effects. Such $\beta$ values would therefore be substantially biased, so we are using the \texttt{AUTO} magnitude based $\beta$ measurements here.

\begin{figure*}  
\centering  
\includegraphics[width=1\linewidth]{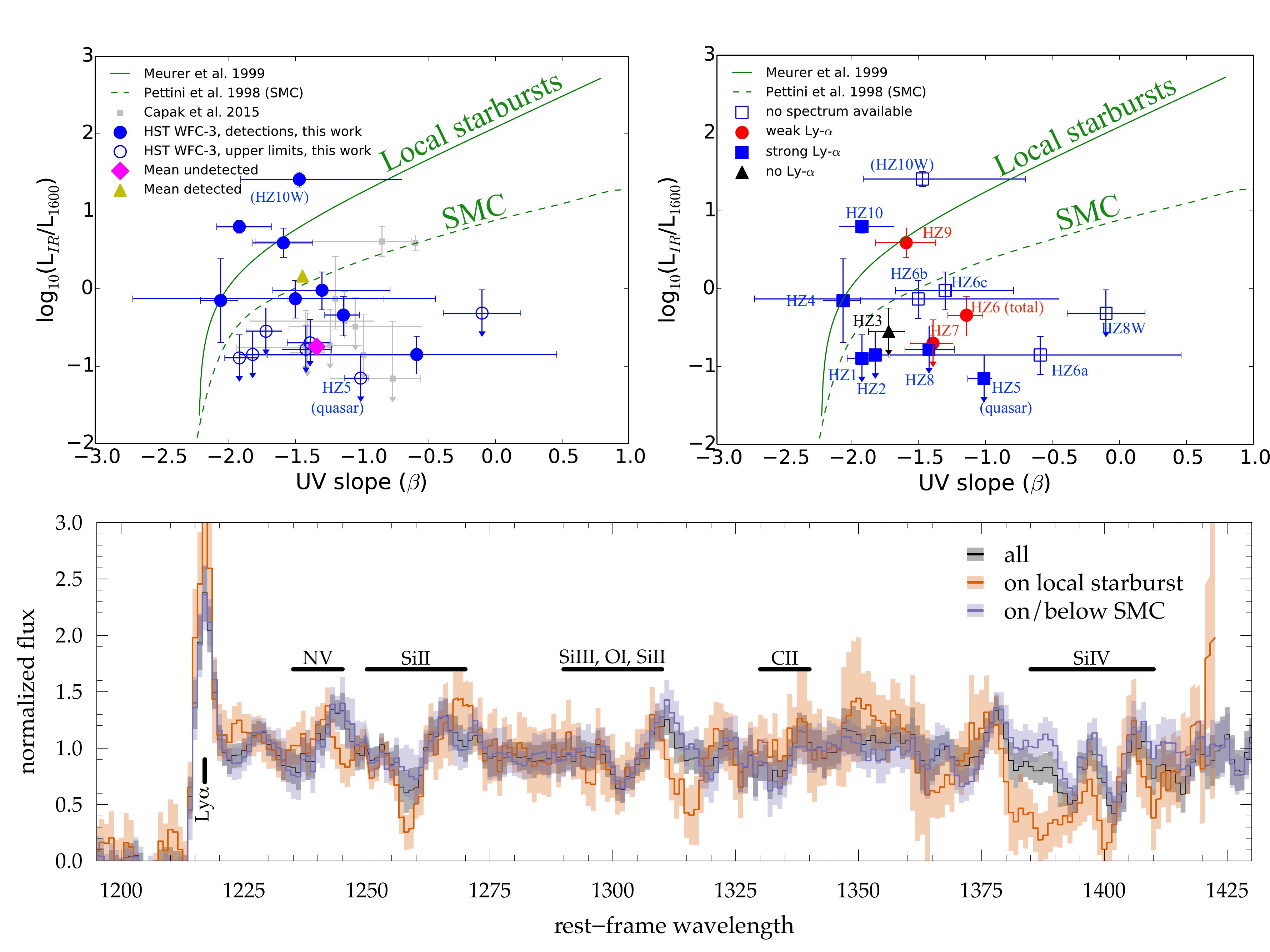}  
\caption{The \IRXB~diagram and spectral properties of our $z\sim5.5$ galaxies.
\textit{Upper left:} The blue symbols show updated $\beta$ and $\LUV$ measurements, and the mean values of detected (undetected) sources are shown with yellow (magenta) symbol. The previous measurement from \citetalias{CAPAK15} are shown by gray symbols. We also show relations from the literature, as indicated in the legend.
\textit{Upper right:} Galaxies color-coded according to their Ly$\alpha$ strength. We do not see a relation between \lya~emission and position on the \IRXB~diagram.
\textit{Bottom:} Stacked Keck/DEIMOS spectra for galaxies falling on the local starburst relation (\textit{HZ4}, \textit{HZ9}, and \textit{HZ10}; red) and below the SMC relation (blue). The black line shows all $9$ galaxies (excluding the low-luminosity quasar HZ5). The $1\sigma$ uncertainties are indicated in colored bands. 
}  
\label{results}  
\end{figure*}  

\section{Results \& Discussion} \label{sec:results} 

	
	The upper left panel of Figure~\ref{results} shows the updated location of the galaxies in blue. The UV luminosities of our study and \citetalias{CAPAK15} are in good agreement within the errors. Our new HST based $\beta$ measurements are consistent within errors with \citetalias{CAPAK15}, but we find a systematic bias of the previous ground-based data toward redder slopes. This is due to the low S/N and systematic uncertainties of that data as noted by \citetalias{CAPAK15} which add a $0.3\,{\rm dex}$ systematic error bar on their measurements.
	Importantly, we have pinned down the $\beta$ of the $z\sim5.5$ galaxies with our new observations with sufficient precision and accuracy to make quantitative statements about the evolution of the \IRXB~relation. We find that roughly $2/3$ of the galaxies are consistent with the dust attenuation models of local starbursts or the SMC extinction law. However, $1/3$ consists of low IRX objects residing \textit{below} the SMC curve with a large range in $\beta$. The redder UV colors could be explained by dust-poor post-starburst galaxies with old stellar populations. However, the Spitzer colors suggesting substantial H$\alpha$ emission, as well as the spectroscopically confirmed strong Ly$\alpha$ line emission, are indicative of young stellar populations for these galaxies (see below). Instead, these galaxies are suggested to have a significant dust attenuation along the line-of-sight, while their low IRX values indicate only very little radiation in the IR, i.e., an overall low dust mass. This is at odds with SMC dust properties as well as other theoretical models that assume dust in thermal equilibrium at $\sim30\,{\rm K}$ that is well mixed with stars \citep[e.g.,][]{CHARLOTFALL00}. 
	To explain these observations, either the physical properties of the dust or the geometry of its spatial distribution have to change. In an upcoming paper, we will investigate this further by investigating in detail the dust temperature and its spatial distribution in high-redshift galaxies (Faisst et al., in prep).
	 
	The $R=3000-4500$ Keck/DEIMOS optical spectra that are available for all 9 main galaxies\footnote{We do not consider the low luminosity quasar \textit{HZ5} here.} allow us to investigate further the spectroscopical properties of the galaxies across the \IRXB~diagram.
	The upper right panel of Figure~\ref{results} shows the \IRXB~diagram with galaxies grouped in strong, weak, and no Ly$\alpha$ emitters. We do not see a trend with Ly$\alpha$ emission strength. 
	The lower panel of Figure~\ref{results} shows median stacks of the spectra of galaxies with dust properties similar to the local starbursts (\textit{HZ4}, \textit{HZ9}, and \textit{HZ10}) and galaxies below the SMC dust extinction curve. The spectra are binned to $1\,{\rm \AA}$ and the $1\sigma$ uncertainties are derived by repeating the stacking analysis 500 times including Gaussian measurement uncertainties.
	This indicates that galaxies similar to local starbursts show deeper absorption in the \siii~and \siiv~doublet ISM lines at $1260\,{\rm \AA}$ and $1400\,{\rm \AA}$, respectively, compared to galaxies below the SMC dust relation. Furthermore, there is indication of a flux deficit around \nv~($\sim1240\,{\rm \AA}$) and an upturn at $1225\,{\rm \AA}$. Such UV absorption features correlate well with the dust column density and gas-phase metallicity of galaxies \citep[e.g.,][]{HECKMAN98,LEITHERER11,FAISST16b}.
	Our findings therefore emphasize the diversity of galaxies at $z\sim5.5$. Galaxies close to the local starburst relation are likely more evolved and metal-rich, while galaxies below the SMC relation that show weaker UV absorption features are dust-poor and metal-poor and possibly of younger age \citep[see also][]{REDDY10}. The latter is also suggested by estimates of the H$\alpha$ equivalent-width from the Spitzer colors of these galaxies (Faisst et al. 2017, in prep), indicating stellar populations of $\textless$ 50Myrs \citep[e.g.][]{COWIE11}.
  Importantly, these young galaxies can have substantially altered spatial dust distributions, which lead to the variety of UV spectral slopes at low IRX value.
	Specifically, clouds of dust may be expelled into the circum-galactic regions of these galaxies by interactions with their environment or internal radiation pressure and turbulences due to vigorous star formation. The serendipitous detections around \textit{HZ8} and the quasar \textit{HZ5} suggest a significant amount of circum-galactic material out to at least $10\,{\rm kpc}$ projected radius. \textit{HZ5} has only a narrow \lya~emission \citep[$30\,{\rm \AA}$ EW and $10\,{\rm \AA}$ FWHM,][]{MALLERY12} and its UV absorption features are at the same redshift as indicated by the \cii~emission of \textit{HZ5a}. This is indicative of the UV undetected \textit{HZ5a} absorbing a substantial part of the quasar's light. 
	\textit{HZ8W} shows one of the reddest $\beta$ but lowest IRX values which could hint toward a several kpc extended foreground dust screen of \textit{HZ8}. \textit{HZ10W} and the components of \textit{HZ6} might be representatives of a similar situation.
	Future follow-up observations of these objects with JWST will be important to further study the inter-galactic medium around these galaxies.

\section{Conclusion} \label{sec:end} 

	We have obtained deep HST near-IR imaging to accurately measure the UV spectral slopes and luminosities of in total $14$ $z\sim5.5$ galaxies. We combine these new measurement with previous ALMA observations at $158\,{\rm \mu m}$ available for all of these galaxies to investigate the \IRXB~diagram at $z\sim5.5$.
	The main findings and results of this letter are as follows:
	
	\begin{itemize}
	
	\item Our new HST observations greatly reduce the uncertainties and biases in UV spectral slopes compared to previous ground-based measurements.  Most of the galaxies at $z\sim5.5$ are either consistent with local starbursts or dust properties similar to the SMC.
	
	\item The galaxies occupy a large range in \IRXB~parameter space, indicative of a large diversity in their properties already 1~billion years after Big Bang.
	
	\item We find galaxies with low IRX values and large ranges of $\beta$, which cannot be explained by SMC-like dust nor models of well mixed stars and dust in thermal equilibrium. Instead the fundamental dust properties or the geometry of dust distribution has to change.
	
	\item Stacked rest-UV spectra of the galaxies across the \IRXB~diagram are indicative of a range of metallicities and evolutionary stages of the galaxies. In particular, galaxies close or below the SMC relation show weak UV absorption features indicative of a low dust and metal content.

	\end{itemize}

\acknowledgements
	We thank the anonymous referee for providing valuable feedback that helped to improve this work.
D.R. and R.P. acknowledge support from the National Science Foundation under grant number AST-1614213 to Cornell University. R.P. acknowledges support through award SOSPA3-008 from the NRAO. V.S. acknowledges support from the European Union'€™s Seventh Frame-work program under grant agreement 337595 (ERC Starting Grant, €'CoSMass'€™).
	Based on observations made with the NASA/ESA Hubble Space Telescope, obtained [from the Data Archive] at the Space Telescope Science Institute, which is operated by the Association of Universities for Research in Astronomy, Inc., under NASA contract NAS 5-26555. These observations are associated with program \#13641 and \#13384.
	Some of the data presented herein were obtained at the W.M. Keck Observatory, which is operated as a scientific partnership among the California Institute of Technology, the University of California and the National Aeronautics and Space Administration. The Observatory was made possible by the generous financial support of the W.M. Keck Foundation. The authors wish to recognize and acknowledge the very significant cultural role and reverence that the summit of Mauna Kea has always had within the indigenous Hawaiian community.  We are most fortunate to have the opportunity to conduct observations from this mountain.
	This paper makes use of the following ALMA data: ADS/JAO.ALMA\#2012.1. 00523.S. ALMA is a partnership of ESO (representing its member states), NSF (USA) and NINS (Japan), together with NRC (Canada), NSC and ASIAA (Taiwan), and KASI (Republic of Korea), in cooperation with the Republic of Chile. The Joint ALMA Observatory is operated by ESO, AUI/NRAO and NAOJ.
	This research has made use of the NASA/IPAC Infrared Science Archive, which is operated by the Jet Propulsion Laboratory, California Institute of Technology, under contract with the National Aeronautics and Space Administration.




\bibliographystyle{aasjournal}
\bibliography{bibtex.bib}





\end{document}